\documentclass[prl,twocolumn,showpacs,preprintnumbers,amsmath,amssymb]{revtex4}
\usepackage{graphicx}

\begin{document}
\title{Controllable $\pi$ junction with magnetic nanostructures}
\author{T. Yamashita,$^{1}$ S. Takahashi,$^{1,2}$ and S. Maekawa$^{1,2}$}
\affiliation{$^{1}$Institute for Materials Research, Tohoku University, 
Sendai, Miyagi, 980-8577, Japan \\
$^{2}$CREST, Japan Science and Technology Agency (JST), 
Kawaguchi, Saitama, 332-0012, Japan}
\date{\today}

\begin{abstract}
We propose a novel Josephson device in which $0$ and $\pi$ states are 
controlled by an electrical current.  
In this system, the $\pi$ state appears in a superconductor/normal metal/superconductor junction 
due to the non-local spin accumulation in the normal metal which is 
induced by spin injection from a ferromagnetic electrode.  
Our proposal offers 
not only new possibilities for application of superconducting spin-electronic devices but also 
the in-depth understanding of the spin-dependent phenomena in magnetic nanostructures.  
\end{abstract}
\maketitle
Nowadays spin-electronics is one of the central topics in condensed matter physics 
\cite{Maekawa1,Maekawa2,Zutic}.  
There has been considerable interest in the spin injection, accumulation, transport, 
and detection in ferromagnet/normal metal (F/N) hybrid structures 
\cite{Johnson,Jedema1,Jedema2,Otani,Son,Takahashi1}.  
Twenty years ago, Johnson and Silsbee demonstrated the spin injection and detection 
in a F/N/F structure for the first time \cite{Johnson}.  
Recently, spin accumulation has been observed at room temperature 
in all-metallic spin-valve geometry 
consisting of a F/N/F junction by Jedema {\it et al.} \cite{Jedema1}.  
In their system, the spin-polarized bias current is applied at one F/N junction, 
and the voltage is measured at another F/N interface 
for the parallel (P) and antiparallel (AP) alignments of the F's magnetizations.  
They have observed the difference of the non-local voltages between the P and AP alignments 
due to spin accumulation in N.  
Also in a F/I/N/I/F (I indicates an insulator) structure, 
a clear evidence of spin accumulation in N has been shown \cite{Jedema2}.  
In hybrid structures consisting of a ferromagnet and a superconductor (S), 
a suppression of the superconductivity due to spin accumulation in S 
has been studied theoretically and experimentally \cite{Takahashi2,Vasko,Dong}.

Furthermore, ferromagnetic Josephson (S/F/S) junctions have been studied actively 
in recent years \cite{Buzdin,Demler,Sellier,Ryazanov,Kontos,Bauer}.  
In the S/F/S junctions, the pair potential oscillates spatially 
due to the exchange interaction in F \cite{Buzdin,Demler}.  
When the pair potentials in two S's take different sign, 
the direction of the Josephson current is reversed 
compared to that in ordinary Josephson junctions.  
This state is called the $\pi$ state in contrast with the $0$ state 
in ordinary Josephson junctions 
because the current-phase relation of the $\pi$ state 
is shifted by ``$\pi$" compared to that of the $0$ state.  
The observations of the $\pi$ state have been reported in various systems experimentally 
\cite{Sellier,Ryazanov,Kontos,Bauer}.  
The applications of the $\pi$ state to the quantum computing also have been proposed 
\cite{Yamashita1,Yamashita2,Ioffe}.
Another system to realize the $\pi$ state is 
a S/N/S junction with a voltage-control channel \cite{Baselmans1,Baselmans2}.  
In the system, the non-equilibrium electron distribution in N 
induced by the bias voltage plays an important role, 
and the sign reversal of the Josephson critical current as a function of 
the control voltage has been demonstrated \cite{Baselmans1,Baselmans2}.

In this paper, we propose a new Josephson device in which 
the $0$ and $\pi$ states are controlled electrically.  
In this device, spin accumulation is generated in a nonmagnetic metal 
by the spin-polarized bias current flowing into the nonmagnetic metal from a ferromagnet.  
In a metallic Josephson junction consisting of the spin accumulated nonmagnetic metal 
sandwiched by two superconductors, the $\pi$ state appears due to the spin split of 
the electrochemical potential in the nonmagnetic metal.  
The magnitude of spin accumulation is proportional to 
the value of the spin-polarized bias current, and therefore 
the state of the Josephson junction is controlled by the current.  
Our proposal leads to an in-depth understanding of the spin-dependent phenomena 
in magnetic nanostructures as well as 
new possibilities for the application of superconducting spin-electronic devices.

\begin{figure}[b]
\includegraphics[width=0.7\columnwidth]{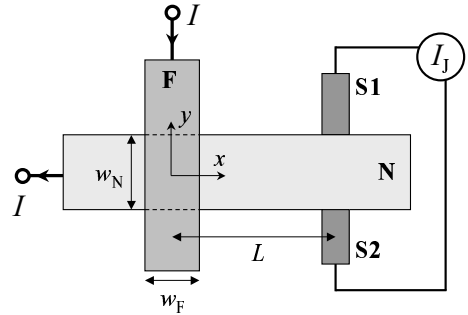}
\caption{Structure of a controllable $\pi$ junction with magnetic nanostructures.  
The bias current $I$ flows from a ferromagnet (F) to the left side of a normal metal (N).  
The Josephson current $I_{\rm J}$ flows 
in a superconductor/normal metal/superconductor (S1/N/S2) junction located at $x=L$.}
\label{Fig1}
\end{figure}
We consider a magnetic nanostructure with two superconductors as shown in Fig. \ref{Fig1}.  
The device consists of a nonmagnetic metal N
(the width $w_{\rm N}$, the thickness $d_{\rm N}$) which is connected to 
a ferromagnetic metal F (the width $w_{\rm F}$, the thickness $d_{\rm F}$) at $x=0$ and 
sandwiched by two superconductors S1, S2 located at $x=L$.  
In this device, the electrode F plays a role as a spin-injector to the electrode N, 
and the S1/N/S2 junction is a metallic Josephson junction.  
The spin-diffusion length $\lambda_{\rm N}$ in N is much longer than 
the length $\lambda_{\rm F}$ in F \cite{Johnson,Jedema1,Jedema2,Otani,Takahashi1}, 
and we consider the structure with dimensions of 
$\lambda_{\rm F} \ll (w_{\rm N(F)}, d_{\rm N(F)}) \ll \lambda_{\rm N}$ which is 
a realistic geometry \cite{Jedema1,Jedema2}.

In the electrodes N and F, the electrical current with spin $\sigma$ is expressed as 
\begin{eqnarray}
{\bf j}_\sigma = -\left({\sigma_\sigma/e}\right) \nabla \mu_\sigma, 
\label{j}
\end{eqnarray}
where $\sigma_\sigma$ and $\mu_\sigma$ are the electrical conductivity and 
the electrochemical potential (ECP) for spin $\sigma$, respectively.  
Here ECP is defined as $\mu_\sigma = \epsilon_\sigma + e\phi$, where 
$\epsilon_\sigma$ is the chemical potential of electrons with spin $\sigma$ and 
$\phi$ is the electric potential.  
From the continuity equation for charge, 
$\nabla\cdot\left({{\bf j}_\uparrow + {\bf j}_\downarrow}\right) = 0$, and that for spin, 
$\nabla\cdot\left({{\bf j}_\uparrow - {\bf j}_\downarrow}\right) 
= e\partial \left({n_\uparrow - n_\downarrow}\right)/\partial t$ 
($n_\sigma$ is the carrier density for spin $\sigma$), we obtain \cite{Takahashi1,Son}
\begin{eqnarray}
&&\nabla^2 \left( {\sigma_\uparrow \mu_\uparrow + \sigma_\downarrow \mu_\downarrow} \right) = 0, 
\label{chem-eqs1}\\
&&\nabla^2 \left( {\mu_\uparrow - \mu_\downarrow} \right) 
= \left( {\mu_\uparrow - \mu_\downarrow} \right)/\lambda^2, 
\label{chem-eqs2}
\end{eqnarray}
where $\lambda = \sqrt{D\tau_{sf}}$ is the spin diffusion length with 
the diffusion constant $D = \left({N_\uparrow + N_\downarrow}\right)/
(N_\uparrow D_\downarrow^{-1} + N_\downarrow D_\uparrow^{-1})$ 
($N_\sigma$ and $D_\sigma$ are 
the density of states and the diffusion constant for spin $\sigma$, respectively) and 
the scattering time of an electron 
$\tau_{sf} = 2/(\tau_{\uparrow\downarrow}^{-1} + \tau_{\downarrow\uparrow}^{-1})$ 
($\tau_{\sigma \bar \sigma}$ is the scattering time of 
an electron from spin $\sigma$ to $\bar \sigma$).  
In order to derive Eqs. (\ref{chem-eqs1}) and (\ref{chem-eqs2}), 
we take the relaxation-time approximation for the carrier density, 
$\partial n_\sigma/\partial t = -\delta n_\sigma/\tau_{\sigma \bar \sigma}$, and 
use the relations $\sigma_\sigma = e^2 N_\sigma D_\sigma$ and 
$\delta n_\sigma = N_\sigma \delta \epsilon_\sigma$, where 
$\delta n_\sigma$ and $\delta \epsilon_\sigma$ are 
the carrier density deviation from equilibrium and 
the shift in the chemical potential from its equilibrium value for spin $\sigma$, respectively.  
In addition, the detailed balance equation 
$N_{\uparrow}\tau_{\uparrow\downarrow}^{-1} = N_{\downarrow}\tau_{\downarrow\uparrow}^{-1}$ 
is also used.  We use the notations 
$\sigma_{\rm N} = 2\sigma_{\rm N}^\uparrow = 2\sigma_{\rm N}^\downarrow$ in N 
and $\sigma_{\rm F} = \sigma_{\rm F}^\uparrow + \sigma_{\rm F}^\downarrow$ 
($\sigma_{\rm F}^\uparrow \neq \sigma_{\rm F}^\downarrow$) in F hereafter.

At the interface between N and F, the interfacial current $I_\sigma$ flows 
due to the difference of ECPs in N and F: 
$I_\sigma = (G_\sigma/e)(\mu_{\rm F}^\sigma |_{z=0^+} - \mu_{\rm N}^\sigma |_{z=0^-})$, where 
$G_\sigma$ is the spin-dependent interfacial conductance.  
We define the interfacial charge and spin currents as 
$I = I_\uparrow + I_\downarrow$ and $I_{\rm spin} = I_\uparrow - I_\downarrow$, respectively.  
The spin-flip effect at the interface is neglected for simplicity.  
In the electrode N with the thickness and the contact dimensions being much smaller than 
the spin-diffusion length ($d_{\rm N}, w_{\rm N}, w_{\rm F} \ll \lambda_{\rm N}$), 
$\mu_{\rm N}^\sigma$ varies only in the $x$ direction \cite{Takahashi1}.  
The charge and spin current densities in N, 
$j = j_\uparrow + j_\downarrow$ and $j_{\rm spin} = j_\uparrow - j_\downarrow$, 
are derived from Eqs. (\ref{j})$-$(\ref{chem-eqs2}), and 
satisfy the continuity conditions at the interface: 
$j = I/A_{\rm N}$ and $j_{\rm spin} = I_{\rm spin}/A_{\rm N}$, where 
$A_{\rm N} = w_{\rm N}d_{\rm N}$ is the cross-sectional area of N.  
From these conditions, we obtain ECP in N, 
$\mu_{\rm N}^\sigma (x) = \overline\mu_{\rm N} + \sigma\delta\mu_{\rm N}$, where 
$\overline\mu_{\rm N} = (eI/\sigma_{\rm N}A_{\rm N})x$ for $x<0$, 
$\overline\mu_{\rm N} = 0$ for $x>0$, and $\delta\mu_{\rm N} = 
(e\lambda_{\rm N}I_{\rm spin}/2\sigma_{\rm N}A_{\rm N})e^{-\left| x \right|/\lambda_{\rm N}}$.  
In the electrode F, the spin split of ECP, $\delta\mu_{\rm F}^\sigma$, 
decays in the $z$-direction because the thickness of F and the dimension of the interface are 
much larger than the spin-diffusion length in F 
($d_{\rm F}, w_{\rm N}, w_{\rm F} \gg \lambda_{\rm F}$) \cite{Takahashi1}.  
In a similar way to the case of N, ECP in F 
is obtained from the continuity conditions for charge and spin currents.  
ECP in F is expressed as 
$\mu_{\rm F}^\sigma (z) = \overline\mu_{\rm F} + \sigma\delta\mu_{\rm F}^\sigma$, where 
$\overline\mu_{\rm F} = (eI/\sigma_{\rm F}A_{\rm J})z + eV$ and 
$\delta\mu_{\rm F}^\sigma = 
(e\lambda_{\rm F}(p_{\rm F}I - I_{\rm spin})/2\sigma_{\rm F}^\sigma A_{\rm J})
e^{-z/\lambda_{\rm F}}$ with 
the contact area $A_{\rm J} = w_{\rm N}w_{\rm F}$, 
the voltage drop at the interface $V = (\overline\mu_{\rm F} - \overline\mu_{\rm N})/e$, and 
the polarization of the current in F, 
$p_{\rm F} = (\sigma_{\rm F}^\uparrow - \sigma_{\rm F}^\downarrow)/\sigma_{\rm F}$.  
The influence of the electrodes S1 and S2 on ECP in N may be neglected.  
When the superconducting gap in S1 and S2 is much larger than 
the spin split $\delta\mu_{\rm N}$ at $x=L$, 
almost no quasiparticle is excited above the gap at low temperature.  
Therefore, the spin current does not flow into S1 and S2, and 
the behavior of ECP in N is not modified by 
the connection to the electrodes S1 and S2.

\begin{figure}[t]
\includegraphics[width=0.7\columnwidth]{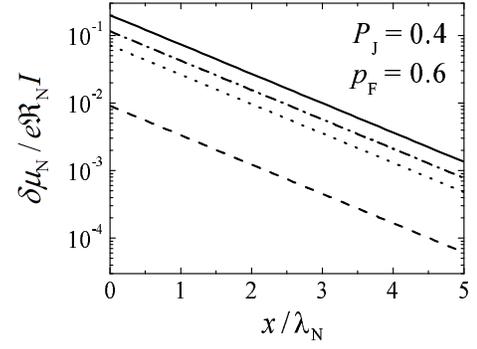}
\caption{Spatial variation of the split of the electrochemical potential in N.  
The solid line is for the tunnel-limit case ($R \gg \Re_{\rm N}, \Re_{\rm F}$), 
the dashed, dotted, and dot-dashed lines are for the metallic-limit cases ($R = 0$) with 
$r = \Re_{\rm F}/\Re_{\rm N} = 0.01$, $0.1$, and $0.2$, respectively.}
\label{Fig2}
\end{figure}
In order to obtain the relation between the bias current $I$ and 
the shift of ECP, $\delta\mu_{\rm N}$, at the right side in N ($x>0$), 
we substitute the obtained $\mu_{\rm N}^\sigma$ and $\mu_{\rm F}^\sigma$ for 
the expressions of $I$ and $I_{\rm spin}$, and eliminate $V$.  
As a result, we obtain the relation between $I$ and $I_{\rm spin}$, and 
finally we get the relation between $I$ and $\delta\mu_{\rm N}$ as follows: 
\begin{align}
&\delta \mu _N \left( x \right) = \nonumber\\
&e\Re_{\rm N} I
\frac{\displaystyle\frac{P_{\rm J}}{1 - P_{\rm J}^2} \left( {\frac{R}{\Re_{\rm N}} } \right) 
+ \frac{p_{\rm F}}{1 - p_{\rm F}^2} \left( {\frac{\Re_{\rm F}}{\Re_{\rm N}}} \right)}
{\displaystyle 1 + \frac{2}{1 - P_{\rm J}^2}\left( {\frac{R}{\Re_{\rm N}}} \right) 
+ \frac{2}{1 - p_{\rm F}^2}\left( {\frac{\Re_{\rm F}}{\Re_{\rm N}}} \right)} 
e^{-x/\lambda_{\rm N}}, 
\label{dm-I}
\end{align}
where $\Re_{\rm N}=\lambda_{\rm N}/(\sigma_{\rm N} A_{\rm N}$) and 
$\Re_{\rm F}=\lambda_{\rm F}/(\sigma_{\rm F} A_{\rm J})$ 
indicate the non-equilibrium resistances of N and F, respectively, 
$R = G^{-1} = {(G_\uparrow + G_\downarrow)}^{-1}$ is the interfacial resistance, and 
$P_{\rm J}=(G_\uparrow - G_\downarrow)/G$ is the polarization of the interfacial current.  
When the F/N interface is the tunnel junction ($R \gg \Re_{\rm N}, \Re_{\rm F}$), 
Eq. (\ref{dm-I}) reduces to a simple form 
$\delta \mu_{\rm N} \left( x \right) = (e\Re_{\rm N}I P_{\rm J}/2) e^{-x/\lambda_{\rm N}}$.  
On the other hand, when the F/N junction is of metallic contact ($R=0$), 
Eq. (\ref{dm-I}) becomes 
$\delta \mu_{\rm N} \left( x \right) = 
e\Re_{\rm N}Ip_{\rm F}re^{-x/\lambda_{\rm N}}/(2r+(1-p_{\rm F}^2))$, where 
$r = \Re_{\rm F}/\Re_{\rm N}$ is a mismatch factor of the resistances in F and N.  
Figure \ref{Fig2} shows the spacial variation of $\delta \mu_{\rm N} \left( x \right)$ 
both for the tunnel- and metallic-limit cases 
with $P_{\rm J}=0.4$ and $p_{\rm F}=0.6$ \cite{Maekawa1,Bass}.  
As shown in this figure, in the case of the metallic contact, 
$\delta \mu_{\rm N}$ becomes larger with decreasing the resistance mismatch \cite{Takahashi1}.

\begin{figure}[t]
\includegraphics[width=0.7\columnwidth]{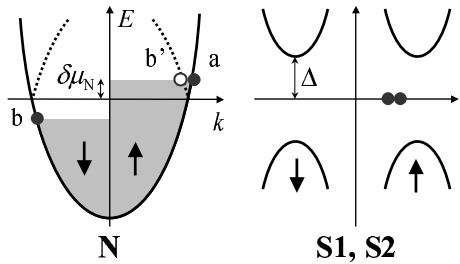}
\caption{Schematic diagram of energy vs. momentum in the Andreev reflection when 
there is spin accumulation in N.  The filled and open circles represent 
an electron and a hole, respectively.  In N, the solid and dashed lines denote 
electron and hole bands, respectively, the shaded area indicates an occupation by electrons.  
In the Andreev reflection, a spin-up electron (a) injected into S's captures 
another electron with spin down (b), and a spin-up hole (b') is reflected back to N.}
\label{Fig3}
\end{figure}
Next we consider how spin accumulation affects the Josephson current $I_{\rm J}$
flowing through the S1/N/S2 junction located at $x=L$ (Fig. \ref{Fig1}).  
In the metallic Josephson junction, 
the Andreev bound state plays a key role for the Josephson effect \cite{Sellier,Andreev}.  
The Andreev bound state is formed by a multiple Andreev reflection 
of an electron with the wave number $k_{e} = (\sqrt{2m}/\hbar)\sqrt{E_F + E}$ and a hole with 
$k_{h} = (\sqrt{2m}/\hbar)\sqrt{E_F - E}$, respectively, where 
$E$ is the energy of the electron and hole measured from the Fermi energy $E_F$.  
As shown in Fig. \ref{Fig3}, when there is the spin split 
$\delta\mu_{\rm N}$ in N, a spin-up (-down) electron with 
the energy $E \approx \delta\mu_{\rm N}$ ($-\delta\mu_{\rm N}$) is 
injected into S's from N at low temperatures.  The injected electron 
captures another electron with the energy $E \approx -\delta\mu_{\rm N}$ ($\delta\mu_{\rm N}$) 
from the opposite spin band in order to form a Cooper pair in S's.  
Therefore, a spin-up (-down) hole with energy 
$E \approx \delta\mu_{\rm N}$ ($-\delta\mu_{\rm N}$) 
is reflected back to N (Andreev reflection \cite{Andreev}).  
In other words, the spin-up (-down) electron with 
$k_{e} \approx (\sqrt{2m}/\hbar)\sqrt{E_F +(-) \delta\mu_{\rm N}}$ and 
the spin-up (-down) hole with 
$k_{h} \approx (\sqrt{2m}/\hbar)\sqrt{E_F -(+) \delta\mu_{\rm N}}$ mainly contribute to 
the formation of a Cooper pair.  
{\it Note that the values of the wave numbers $k_{e}$ and $k_{h}$ differ due to 
the spin split $\delta\mu_{\rm N}$ in contrast with 
the case of no spin split {\rm(}$\delta\mu_{\rm N}=0${\rm)} in which $k_{e} \approx k_{h}$}.

The split $\delta\mu_{\rm N}$ corresponds to the exchange energy $E_{ex}$ of a ferromagnet 
in a superconductor/ferromagnet/superconductor (S/F/S) Josephson junction 
as follows \cite{Buzdin,Demler,Sellier,Ryazanov,Kontos,Bauer}: 
In the S/F/S systems, 
Cooper pairs are formed by the Andreev reflection of 
spin-$\sigma$ electrons with the wave number 
$k_{e,\sigma}^{\rm F} \approx (\sqrt{2m}/\hbar)\sqrt{E_F + \sigma E_{ex}}$ and holes with 
$k_{h,\sigma}^{\rm F} \approx (\sqrt{2m}/\hbar)\sqrt{E_F - \sigma E_{ex}}$ 
at the energy $E \approx 0$.  
In the case that the exchange interaction is much weaker than 
the Fermi energy ($E_{ex} \ll E_F$), the stable state ($0$ or $\pi$) in the system 
depends on the dimensionless parameter $\alpha_{\rm F} = (E_{ex}/E_F)(k_F d_{\rm F})$, 
where $d_{\rm F}$ is the thickness of F and $k_F$ is the Fermi wave number \cite{Sellier}.  
At $\alpha_{\rm F} = 0$ the system is in the $0$ state, and 
the first $0$-$\pi$ transition occurs at $\alpha_{\rm F} = \pi/2$, and then 
the system is in the $\pi$ state at $\alpha_{\rm F} = \pi$ \cite{Sellier}.  
Because the value of $E_{ex}$ is fixed in the S/F/S system, the $0$ and $\pi$ states change 
periodically with the period $2\pi(E_F/E_{ex})$ as a function of $d_{\rm F}$.  
As a result, the $d_{\rm F}$ dependence of the Josephson critical current shows 
a cusp structure and the critical current becomes minimum 
at the $0$-$\pi$ transition \cite{Kontos,Sellier}.

In analogy with the case of the S/F/S junction discussed above, 
when there is spin accumulation in N as shown in Fig. \ref{Fig3}, 
the $0$ or $\pi$ state is realized in the S1/N/S2 junction depending on the parameter 
$\alpha = (\delta\mu_{\rm N}/E_F)(k_F w_{\rm N})$.  
In this case, the width $w_{\rm N}$ is fixed, and 
the $0$ and $\pi$ states are controlled through the value of $\delta\mu_{\rm N}$ 
which is proportional to the bias current $I$ (see Eq. (\ref{dm-I})).  
The N part of the system is in the non-equilibrium state by the spin current 
in contrast with F in the equilibrium state of the S/F/S junction.  
However, one can discuss the critical current in the non-equilibrium S1/N/S2 junction 
in the same way as the equilibrium S/F/S junction 
because the critical current is dominated by the energy of the quasiparticles in N, 
not by the flow of the current \cite{Baselmans1,Baselmans2}.

From the point of view of more detailed description, 
the free energy in the system is obtained by the summation of 
the energy of the Andreev bound states \cite{Yamashita1}.  
The bound state energy is calculated from the Bogoliubov-de Gennes equation \cite{BdG}, 
and the free energy is minimum for the phase difference $0$ ($\pi$) for the $0$ ($\pi$) state.  
In the S1/N/S2 junction with no spin accumulation in N ($\delta\mu_{\rm N}=0$), 
the bound states with the energy $E>0$ contribute to the free energy.  
On the other hand, when spin accumulation exists in N, 
the spin-up (-down) bound states with the energy $E>\delta\mu_{\rm N}$ ($-\delta\mu_{\rm N}$) 
contribute to the free energy because ECP is shifted by 
$\delta\mu_{\rm N}$ ($-\delta\mu_{\rm N}$) in N.  
The $0$-$\pi$ transition occurs due to the shift of 
the energy region of the Andreev bound states which contribute to the free energy.

As an example, we consider the case that the F/N interface consists of a tunnel junction.  
The material parameters 
$P_{\rm J}=0.4$, $\rho_{\rm N}=\sigma_{\rm N}^{-1}=2\,\mu\Omega{\rm cm}$, 
$\lambda_{\rm N}=1\,{\rm \mu m}$, $w_{\rm N}=800\,{\rm nm}$, and $d_{\rm N}=10\,{\rm nm}$, 
which lead to $\Re_{\rm N}=2.5\,\Omega$, are taken.  
The distance between F and S's is taken to be $L = 500\,{\rm nm}$.  
When no bias current is applied between F and N ($I=0$), the S1/N/S2 junction is 
in the ordinary $0$ state because there is no spin split of 
ECP ($\delta\mu_{\rm N} = 0$).  
With increasing the bias current, the magnitude of 
the Josephson critical current decreases because the parameter $\alpha$ increases 
due to the increase of the spin split.  
When the bias current reaches the value $I = I_0 \approx 3\,{\rm mA}$ which induces 
the spin split $\delta\mu_{\rm N} \approx 1\,{\rm meV}$ at $x=500\,{\rm nm}$, 
the parameter $\alpha \approx \pi/2$ and 
the first transition to the $\pi$ state from the $0$ state occurs 
(the values of $E_F = 5\,{\rm eV}$ and $k_F = 1\,{\rm \r{ A}^{-1}}$ are taken \cite{AM}).  
As a result, the magnitude of the Josephson critical current takes its minimum 
at $I = I_0$, and increases with increasing the bias current $I > I_0$.  
When the bias current attains $I = 2I_0$, the magnitude of the Josephson critical current 
becomes maximum because of $\alpha \approx \pi$, 
and decreases with increasing the bias current $I > 2I_0$.  
For $I = 3I_0$ corresponding to $\alpha \approx 3\pi/2$, 
the second transition to the $0$ state from the $\pi$ state occurs.

Here we discuss the effect of spin accumulation on the superconducting gap \cite{Takahashi2}.  
The spin split $\delta\mu_{\rm N}$ at $x=L$ in N causes the split of 
ECP of S's by $\delta\mu_{\rm N}$ near the S/N interfaces.  
The spin split in S's decreases exponentially 
with the spin-diffusion length $\lambda_{\rm S}$ from the interface.  
In the superconductors, the superconducting gap is not suppressed by spin accumulation 
until $\delta\mu_{\rm N}$ exceeds the critical value of 
the spin split $\delta\mu_{{\rm N}c}$ \cite{Takahashi2}.  
At low temperatures much lower than the superconducting critical temperature ($T \ll T_c$), 
the critical value of the spin split is obtained as $\delta\mu_{{\rm N}c} \lesssim \Delta_0$ 
by solving the gap equation \cite{Takahashi2}, 
where $\Delta_0$ is the superconducting gap for $\delta\mu_{\rm N}=0$ at $T=0$.  
In the case discussed in the above paragraph, $\delta\mu_{\rm N} \approx 1\,{\rm meV}$ at 
the first $0$-$\pi$ transition ($\alpha \approx \pi/2$).  
For example, $\Delta_0 \approx 1.5\,{\rm meV}$ for niobium \cite{Kittel}, 
and therefore the superconducting gap is almost not affected 
by spin accumulation at the first $0$-$\pi$ transition.  
When superconductors with the higher value of $T_c$, 
e.g., ${\rm MgB}_2$ ($T_c$ $\approx$ 39 K) \cite{Nagamatsu} or 
High-$T_c$ materials ($T_c$ is several 10 K's) \cite{Kittel}, 
are used as the electrodes S1 and S2, the superconductivity is robust 
even at the second ($\delta\mu_{\rm N} \approx 3\,{\rm meV}$, $\alpha \approx 3\pi/2$) 
and higher $0$-$\pi$ transitions.

In summary, we have proposed the novel Josephson device in which 
the $0$ and $\pi$ states are controlled electrically.  
The spin split of the electrochemical potential is induced in the electrode N 
by the spin-polarized bias current flowing from F to N.  
The $\pi$ state appears in the S1/N/S2 junction due to the non-local spin accumulation in N.  
Because the magnitude of spin accumulation 
is proportional to the value of the spin-polarized bias current, 
the $0$ and $\pi$ states of the Josephson junction are controlled by the current.  
Our proposal provides 
not only new possibilities for the application of superconducting spin-electronic devices 
but also the deeper understanding of the spin-dependent phenomena in the magnetic nanostructures.

We are grateful to M. Mori and G. Montambaux for fruitful discussion.  
T.Y. was supported by JSPS Research Fellowships for Young Scientists.  
This work was supported by NAREGI Nanoscience Project, Ministry of 
Education, Culture, Sports, Science and Technology (MEXT) of Japan, 
and by a Grant-in-Aid from MEXT and NEDO of Japan.

\end{document}